# Discovering Communities of Malapps on Android-based Mobile Cyber-physical Systems

Dan Su, Jiqiang Liu, Wei Wang*, Xiaoyang Wang, Xiaojiang Du, Mohsen Guizani

*Abstract*—Android-based devices like smartphones have become ideal mobile cyber-physical systems (MCPS) due to their powerful processors and variety of sensors. In recent years, an explosively and continuously growing number of malicious applications (malapps) have posed a great threat to Android-based MCPS as well as users' privacy. The effective detection of malapps is an emerging yet crucial task. How to establish relationships among malapps, discover their potential communities, and explore their evolution process has become a challenging issue in effective detection of malapps. To deal with this issue, in this work, we are motivated to propose an automated community detection method for Android malapps by building a relation graph based on their static features. First, we construct a large feature set to profile the behaviors of malapps. Second, we propose an E-N algorithm by combining epsilon graph and k-nearest neighbor (k-NN) graph for graph construction. It solves the problem of an incomplete graph led by epsilon method and the problem of noise generated by k-NN graph. Finally, a community detection method, Infomap, is employed to explore the underlying structures of the relation graph, and obtain the communities of malapps. We evaluate our community detection method with 3996 malapp samples. Extensive experimental results show that our method outperforms the traditional clustering methods and achieves the best performance with rand statistic of 94.93% and accuracy of 79.53%.

## I. INTRODUCTION

The popularity of mobile devices has increased great interest in the area of mobile cyber-physical systems (MCPS). Smartphones have become ideal MCPS due to their powerful processors and variety of sensors. Android has been the most popular mobile operating system for smartphones with a global market share of 87.6% [1]. However, the open nature of Android system and the readily-available application distribution mechanism attract both developers and attackers. Malapps are developed to steal personal information or gain control of devices illegally. More severely, the attack can easily expend from the users' cyber world to physical world, which will pose a great threat to users' privacy and properties. Symantec [2] indicated that the volume of new Android malapp variants was up to 3.6 thousand. They blocked 18.4 million mobile malapp infections in 2016. The significant growth of malapps targeting Android-based MCPS increasingly requires efficient methods that can automatically profile and categorize malapps.

In order to cut costs and accelerate the development process, attackers tend to inject malicious components into an existing malapp and publish it after reassembling. Since code reuse is widely adopted in malapp programming, though large amount of malapps arise every day, most of them are variants of existing malapp families [3]. Samples in the same family share similar behaviors and show similar vulnerabilities. Malapp family classification can filter out existing malapp variants quickly and leave the others for further analysis. Hence it can expedite malapp detection process. In addition, exploring the relations among malapps and studying the evolution process of families can help to get a better understanding of malapp development, and further forecast the developing trend. Therefore, categorizing malapps is very helpful for maintaining the security of Android ecosystem.

Although there exists work focusing on characterizing malapp families, some problems remain unsolved. First, the boundaries of malapp families are difficult to define. Some behaviors are common among malapps, such as connecting the Internet or sending phone identifiers to remote servers. These similar behaviors will result in the similarity of different families and the obscurity of their boundaries. How to precisely characterize malapps and find the families' patterns remain unaddressed. Second, most existing work on categorizing malapp families is based on supervised machine-learning classification methods. The high precision achieved by some methods depend on the large labelled dataset in the training process. The requirement of labelled data, however, limits the ability to detect new malapp families. On the other hand, for the traditional unsupervised clustering methods, e.g., k-means, the evaluation of similarities among malapps in which is too coarse to capture the implicit information and the underlying relations of malapps.

As new malapps appear very frequently, alternative methods that are able to discover novel malapp families have become a necessity rather than an option. In this work, we treat malapps

The work reported in this paper was supported in part by National Key R&D Program of China, under grant 2017YFB0802805, in part by the Scientific Research Foundation through the Returned Overseas Chinese Scholars, Ministry of Education of China, under Grant K14C300020, in part by the Fundamental Research Funds for the Central Universities of China, under grant K17JB00060, K17JB00020 and KKJB17019536, and in part by Natural Science Foundation of China, under Grant U1736114 and 61672092.

D. Su, J. Liu, W. Wang are with Beijing Key Laboratory of Security and Privacy in Intelligent Transportation, Beijing Jiaotong University, Beijing 100044, China.(Email: sudan1@bjtu.edu.cn; jqliu@bjtu.edu.cn; wangwei1@bjtu.edu.cn).

X. Wang is with Beijing Key Laboratory of Traffic Data Analysis and Mining, Beijing Jiaotong University, Beijing 100044, China. (Email: shawnwang@bjtu.edu.cn).

X. Du is with Dept. of Computer and Information Sciences, Temple University, Philadelphia PA 19122, USA. (Email: dxj@ieee.org).

M. Guizani is with Dept. of Electrical and Computer Engineering, University of Idaho, Moscow, Idaho, USA. (Email: mguizani@ieee.org).

in a correlation perspective. The relationship between malapps and their associated families is like individuals and organizations. Since individuals in the same organization are related to each other by various interdependencies, the more similar the samples' behaviors are, the more likely they are in the same organization. Community detection techniques can be leveraged to discover relations between interacting individuals. For example, in a social network, people in the same community share the same schools or hobbies. In a biological protein network, communities are functional modules of interacting proteins [4]. Similarly, in a malapp relation graph, malapps can be regarded as vertices and the relationships can be represented by weighted edges. Malapps that behave similarly would gather into the same community. The newly arising malapps will deviate from existing families and form their own groups. Thus both variants from known families and novel malapp families can be discovered.

To deal with the security issue on Android-based MCPS, in this paper, we propose a framework for malapp classification based on the community concept. The main idea is to build a relation graph for malapps and apply community detection methods for community discovery. First, we disassemble the Android Package (APK) file of each malapp and extract information to characterize its behaviors. The features fall into 11 categories, which profile each malapp in different aspects. Second, we calculate the weights between each pair of malapps based on the features. The weights represent similarities among malapps. Variants in the same family tend to have higher weights because of their similar behaviors. Third, we generate an undirected weighted graph for all the malapps based on the weights. We name this graph relation graph since it represents the relations among malapps. Finally, we employ Infomap, a community detection algorithm, to extract the highly-inner-connected communities in the graph. Community detection methods are able to better explore the implicit similarities of interrelated data than traditional clustering methods. Our approach does not need to know any prior patterns of malapps. It is able to classify not only variants of existing malapp families, but also newly arising families.

We make the following contributions:

(1) We construct a large feature set that contains 11 categories to characterize the behaviors of malapps on Android-based MCPS. According to the generality, the features of 11 categories are divided into two types, app-specific and platform-defined. We evaluate different feature sets in terms of their effectiveness and limitations on the detection of malapps.

(2) We propose a graph construction algorithm named E-N method. The commonly used graphs, epsilon graph and k-NN graph suffer from isolated vertices and over-balanced problems, which could result in an incomplete graph and much noise during the process of graph construction. We combine the two graphs together to overcome these disadvantages. The empirical results show that the E-N method achieves better efficiency and accuracy.

(3) We apply community detection techniques to group malapps and uncover their underlying relations. Community detection methods are rarely applied in the area of exploring the relations of Android malapps. They can explore underlying relations of malapps and show better adaptability to noise and outliers than traditional clustering methods. We provide detailed explanations about their superiority in malapp detection.

(4) We conduct a series of experiments to evaluate our community detection method based on E-N algorithm. The experimental results demonstrate that our method outperforms the traditional clustering methods and achieves the best performance with rand statistic of 94.93% and accuracy of 79.53%.

The rest of this paper is organized as follows. Section II introduces related work on Android malapp detection and family classification. Section III describes the proposed method. Section IV describes the experiments and discussion. The conclusion follows in Section V.

## II. RELATED WORK

Several approaches addressing the security issue of Android-based MCPS have been presented in the literature. For detecting and classifying malapps, static analysis and dynamic analysis are normally applied for malapp detection.

Static features are extracted from decompiled APK files. Arp et al. [6] presented a system called DREBIN, which extracted static features such as component names, permissions, intents, API, etc. They considered linear SVM for the training task, and the comprehensive feature set at that time inspired later researchers. Shabtai et al. [7] extracted features from Android app files, such as Java bytecode and XML files. They tested several feature selection approaches to find the most representative sets of features. McLaughlin et al. [8] used a deep convolutional neural network to learn n-gram features of malapps' opcode sequence. Andow et al. [9] proposed an automated approach for resolving the semantics of user inputs requested by mobile applications. They identified some concerning practices, such as insecure exposure of account passwords and achieved an overall accuracy of 95% at semantics resolution. In our previous work [10], we ranked the permissions according to their risks and systematically studied how well permissions perform in the detection of malapps. We built a framework [11] to effectively manage a big app market in terms of detecting malapps and categorizing benign apps. We also developed a static analysis tool called SDFDroid [12] that identified the sensors' types and generated the sensor data propagation paths in each app to provide an overview of sensor usage patterns. To detect novel malapps, we developed an anomaly detection system called Anomadroid [13] to profile normal behaviors of normal apps. Apps whose behaviors deviated from the normal profile were identified as malicious.

In general, dynamic analysis executes an app in the emulator and monitors its behaviors in the runtime. Enck et al. [14] proposed TaintDroid to figure out how apps collected and shared private data with dynamic taint analysis. It tracked multiple sources of sensitive data and monitored their sinks. Zhao et al. [15] introduced AMDetector which employed a hybrid static-dynamic analysis method to build an attack tree. The result showed that the True Positive Rate is 88.14%. A number of researchers paid attention to network flows. Malapps need to connect remote severs to receive commands and send sensitive user data. Http header and statistic information (e.g., number of received packages) [16-21] were collected to profile malapps' network behaviors.

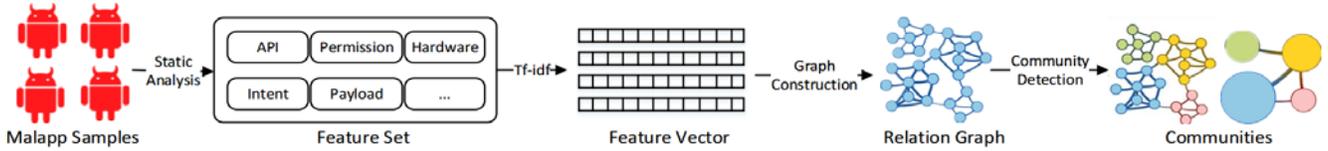

Fig. 1. Framework of our method

Classifying malapps into families can quickly filter out existing malapp variants from massive newly generated malapps and expedite detection process. Suarez-Tangil et al. [22] proposed DENDROID. They adopted the standard Vector Space Model to describe control flow structures and applied k-NN to classify malapp families. Dash et al. [23] developed DroidScribe to focus on app's runtime behaviors such as network access, file access, binder methods, etc. They proposed a prediction technique to improve SVM performance. Hsiao et al. [24] extracted API call sequences and applied phylogenetic tree, PCA and dot matrix to demonstrate malapps' behavioral correlations. Since most of the malapps are repackaged, detection algorithms based on graph similarity are widely adopted. Zhou et al. [25] built sensitive API call graphs and compared their maximum isomorphic subgraphs to classify malapp families. Hu et al. [26] proposed MIGDroid to construct method invocation graphs on the Smali code level, and then divided the method invocation graphs into weakly connected sub-graphs. It calculated threat scores based on sensitive APIs. Subgraphs with higher scores were more likely to be malicious. Crussell et al. [27] proposed DNADroid to detect clone apps by comparing program dependency graphs between methods in candidate applications. Zhang et al. [28] proposed ViewDroid and created a feature view graph based on users' navigation behaviors across app views. It characterized Android apps from a higher-level abstraction, making it resilient to code obfuscation. Key management ([29-31]) is essential for security. Several papers (e.g., [32-35]) have studied related security issues.

Although there exists work focused on malapp family classification, however, some problems remain unsolved. Graph similarity based methods suffer from high computational complexity. In addition, most supervised methods can only classify malapps with known labels and cannot handle new malapp families. Moreover, the traditional unsupervised methods are unable to demarcate families because of the coarse evaluation method. Therefore, to deal with novel malapps and uncover their relations, we apply community detection techniques to explore communities of malapps.

III. METHOD

The framework of our method consists of four phases: feature extraction, data preparation, relation graph construction and community detection. As described in Fig. 1, after the malapps are fed into the framework, a large number of features will be extracted from each APK file. We utilize tf-idf to measure the importance of features. Then we calculate the weights between each pair of malapps based on the features. In the graph construction phase, we propose the E-N algorithm to build a relation graph for malapps based on the weights. Finally, a community detection algorithm is applied to discover the underlying structures of the graph.

A. Feature Extraction

Representative and comprehensive features are in great demand to profile malapps' behaviors. Since dynamic analysis is highly consumptive of system resources, and how to traverse all paths to trigger more behaviors is another difficulty remaining unsolved. In this work, we employ static analysis, so as to reach the goal of being efficient. We extract 11 categories of features from AndroidManifest.xml and Dex files. The large feature set can characterize an app in different aspects. The summary of features can be seen in Table I.

TABLE I FEATURE SET

|  | Feature Name | Data Type | Feature Type |
|---|---|---|---|
| $FS_1$ | Requested permissions | Boolean | Platform-defined/App-specific |
| $FS_2$ | Filtered intents | Numeric | Platform-defined/App-specific |
| $FS_3$ | Restricted API | Numeric | Platform-defined |
| $FS_4$ | Components names | Boolean | App-specific |
| $FS_5$ | Code patterns | Boolean | Platform-defined |
| $FS_6$ | Certification information | Boolean | App-specific |
| $FS_7$ | Payload file types | Numeric | Platform-defined |
| $FS_8$ | Strings | Numeric | App-specific |
| $FS_9$ | Used permissions | Boolean | Platform-defined |
| $FS_{10}$ | Hardware | Boolean | Platform-defined |
| $FS_{11}$ | Suspicious calls | Numeric | Platform-defined |

These static features are divided into two types according to their generality and specificity: platform-defined and app-specific. Platform-defined features are defined by Android system and can be applied to all apps. App-specific features are defined by developers and are sometimes unique to certain apps. In section IV, we make a comparison between the feature sets and discuss their advantages as well as limitations.

The values of features have two types: some features are Boolean, e.g., whether the permission is declared. Others are numeric, e.g., the frequency of a certain API is invoked. We use numeric features instead of binary features because they contain more information. Meanwhile, the effectiveness of tf-idf relies on the computation of frequencies.

B. Data Preparation

High dimensional feature vectors are required for further analysis. A feature vector, which represents a malapp, is defined as $F = (f_1, f_2, ... f_n)$, where $n$ is the number of features and $f_i$ is the value of $i$-th feature. If the malapp does not have the $i$-th feature, $f_i$ would be set to 0.

Term frequency-inverse document frequency (Tf-idf) [36] is a numerical statistic method that measures the importance of a word to the whole text. We employ tf-idf to measure the representativeness of features. It can be defined in logarithmic form as

$$tfidf_{mj} = tf_{mj} \times log\left(\frac{n}{s_m}\right) \quad (1)$$

where $tf_{mj}$ is the frequency of feature $m$ in sample $j$, $n$ is the size of dataset, $s_m$ is the total number of samples that contain feature $m$ in the dataset. If $tfidf_{mj}$ is high, it means feature $m$ appears frequently in sample $j$, but rarely appears in other samples, which makes feature $m$ highly representative of sample $j$.

To quantify the closeness among malapps, we calculate the weights for each malapp pair $i$ and $j$, which is defined as

$$w_{ij} = \sum_k tfidf(f_k) \quad (2)$$

where $f_k$ is the common feature between $i$ and $j$. The weight indicates both the quantity and importance of common features between malapps. In general, if malapp pair $i$ and $j$ share a high weight, they hold high potential to be similar.

### C. Relation Graph Construction

The relation graph construction phase is essential to community detection. A well-defined graph reveals relations among malapps, not only by the directly connected vertices, but also by their neighbors. After phases of feature extraction and data preparation, we get feature vectors that profile malapps' behaviors and weights of malapp pairs which indicate their similarities. Based on these weights, we build a co-occurrence relation graph for all malapps in the dataset. Two samples are more likely to establish a connection if specific features appear in both of them. The relation graph is an undirected weighted graph, which is defined as $G = (V, E, W)$. $V$ is the vertex set and each $v \in V$ represents a malapp. $E$ is the edge set. There is an edge between two vertices if two malapps share a high weight. $W$ is the weight set of the edges and the $w_{ij}$ in $W$ is defined in (2). The most commonly used graphs are the k-nearest neighbor (k-NN) graph and the epsilon graph.

The k-nearest neighbor graph is a graph in which two vertices, $v_i \in V$ and $v_j \in V$, are connected by an edge, if the weight between $v_i$ and $v_j$ is among the $k$ largest weights of $W_i$. In order to find the k-nearest neighbors, weights between $v_i$ and any $v_j \in V$ need to be sorted. The sorting process could be a bottleneck, since the time complexity is too high if the vertex set is large. Furthermore, k-NN over balanced the number of nearest neighbors. If a vertex has high similarities with a quantity of vertices, it would lose lots of similar neighbors by connecting only $k$ of them. On the other hand, a marginal vertex that has low similarity with others would be forced to link to $k$ vertices, which would increase the noise in the relation graph. It further leads to the difficulty of the selection of $k$ since $k \in [1, n]$, where $n$ is the number of vertices. The time complexity and the over balanced problem have become its main drawbacks.

In an epsilon graph, an edge will be created between every vertex pair $v_i$ and $v_j$ if $w_{ij} > \varepsilon$. Compared with k-NN, the construction process is more efficient. Since $\varepsilon$ is a constant and global measurement for all malapps, the degrees of vertices are entirely based on their closeness with others, instead of being set manually. However, the parameter $\varepsilon$ cannot characterize the relations between vertices universally. If $\varepsilon$ is small, numerous vertices that are not so similar with each other will be connected, and a dense complete graph is formed with fuzzy community structures. Thus we cannot achieve a clear observation of the communities' distribution. If we pick a large $\varepsilon$, only the most similar vertices would be joint together. But in the meantime, the number of isolated vertices will increase, which would lead to incompleteness of the graph, and further limit community detection progress.

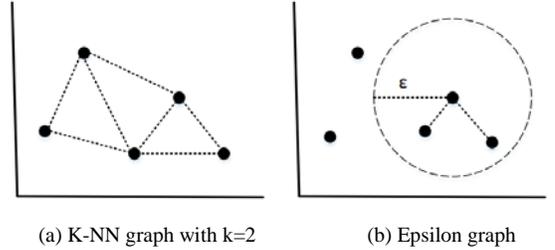

(a) K-NN graph with k=2    (b) Epsilon graph

Fig. 2. An example of k-NN graph and epsilon graph

Since the k-NN graph and the epsilon graph both have disadvantages, to achieve the goal of fast constructing a distinct and representative relation graph of malapps, we propose the E-N method by combining the two graphs together. The graph construction process consists of two steps: Firstly, an edge $e_{ij}$ is created if $w_{ij}$ is among the top $p$ percent of $W$, where $p$ is defined as $p = \frac{|\{w_{ij} \geq \varepsilon\}|}{|W|} \times 100$. Based on our experimental results, when $p$ is around 10, the edges can best demonstrate relations between vertices. Secondly, for the few isolated vertices, k-NN is applied to give connections to their $k$ nearest neighbors. In this case, the disadvantages of over balance and high time complexity in k-NN graph are made up by epsilon graph, and in the meantime, the problem of isolated vertices in epsilon graph is addressed by k-NN. The E-N algorithm achieves complementarity of epsilon method and k-NN method. It is described below.

---
**Algorithm 1: E-N Method**
---
**Input**: $V, W$
1.    **for** each $v_i$ and $v_j$
2.       **if** $w_{ij} > \varepsilon$
3.          create edge $e_{ij}$
4.    **end for**
5.    **for** each $v_m$ that has no edge
6.       find its k-nearest neighbors $neighbours(v_m)$
7.       **for** each $v_n \in neighbours(v_m)$
8.          create edge $e_{mn}$
9.    **end for**

**Output**: $E$

### D. Community Detection

Various malapp families can form inner-dense communities

[37] within the relation graph. Infomap [38] is employed to extract the underlying structures of the graph. We assume that there is a flow of information on the graph moving along the links between graph entities. The graph structure represented by links will influence and control the directions of the flow. If we assign each vertex a unique code, the steps of the flow trace with a series of encoding can be described. Next, we apply a flow-based and information-theoretic method, which is known as a map equation, to quantify how well the structures drive this flow across the graph. For the malapp relation graph, the map equation specifies the theoretical limit of how concisely we can describe the trajectory of the flow on the graph. Minimizing the map equation over all possible graph partitions reveals which structure can most probably lead to such a flow trajectory. Thus the problem of detecting underlying structures can be transformed into a coding compression problem: how to compress the code for describing locations of the flow trace.

The malapp relation graph can be decomposed into a set of communities. Once the flow enters a community, it tends to remain flowing in the community for a long time because intra-community vertices are more densely connected than inter-community vertices. A codeword is a sequence of binary code assigned to a vertex. A set of codewords compose a codebook, which is used to describe vertices in a community. By implementing separate codebooks for different communities, the average code length will be shorter than a global Huffman code. According to Shannon's source coding theorem [38], which implies that when we use $n$ codewords to describe the $n$ states of a random variable $X$ that occur with frequencies $p_i$, the average length of codewords can be no less than the entropy of the random variable $X$ itself: $H(X) = -\sum_1^n p_i log(p_i)$. If we partition the graph with $n$ vertices into $m$ communities, the lower limit of code length can be defined as

$$L(M) = q_{\curvearrowright} H(Q) + \sum_{i=1}^{m} p_{\circlearrowleft}^i H(P^i) \quad (3)$$

Here $H(Q)$ is the entropy of the index codebook. Index codebook is used to switch between communities. $H(P^i)$ is the entropy of codebook $i$. $q_{\curvearrowright}$ is the probability that the flow exits communities and $p_{\circlearrowleft}^i$ is the probability the flow stays in community $i$ and exits the community. The algorithm of Infomap is described below.

---
**Algorithm 2: Infomap**
**Input**: $E, W$
1. assign each $v_i$ with community $c_i$
2. compute map equation $e_{min}$
3. **for** each $v_i$ randomly
4.    **for** each $v_j \in nieghbours(v_i)$
5.       Assume $v_i$ is assigned with community $c_j$
        compute map equation $e_j$ and difference $\Delta e = e_j - e_{min}$
6.    **end for**
7.    **if** $min\Delta e \geq 0$
8.       do nothing
9.    **else**
10.       assign $v_i$ with community $c_j$, $j = \min_j \Delta e$
11. **end for**
12. **repeat** step 2-11 until all vertices' map equation do not decrease
13. rebuild the graph, with the last-level communities forming the vertices at this level
14. **repeat** step 2-13 until the map equation no longer decreases

**Output**: *Communities*

---

### E. Measurement

We apply two measurements to evaluate our community detection method, Rand Statistic (RS) and Accuracy (Acc).

Define $P = \{p_1, p_2, \ldots p_n\}$ as vertices' original families, and $C = \{c_1, c_2, \ldots c_m\}$ as their newly discovered communities in the graph.

(1) Rand Statistic (RS)

We measure the partition of the communities by comparing the relations of each vertex pair $(v_i, v_j)$ in $P$ and $C$. We use $ss$ to denote the case that $v_i$ and $v_j$ are in the same $P$ and also in the same $C$. $N_{ss}$ denotes the quantity of vertex pairs that satisfy the condition of $ss$. Similarly, the meaning of $sd, ds, dd$ can be seen in Table II. Apparently, $N_{ss} + N_{sd} + N_{ds} + N_{dd} = C_n^2$. The definition of Rand Statistic is

$$RS = \frac{N_{ss}+N_{dd}}{N_{ss}+N_{sd}+N_{ds}+N_{dd}} \quad (4)$$

The larger RS is, the more coincident $P$ and $C$ are.

TABLE II NOTATION

| General notation | Meaning |
|---|---|
| $P = \{p_1, p_2, \ldots p_n\}$ | Original families |
| $C = \{c_1, c_2, \ldots c_m\}$ | Discovered communities |
| $ss$ | $v_i$ and $v_j$ are in same $P$ and $C$ |
| $sd$ | $v_i$ and $v_j$ are in same $P$ but different $C$ |
| $ds$ | $v_i$ and $v_j$ are in different $P$ but same $C$ |
| $dd$ | $v_i$ and $v_j$ are in different $P$ and $C$ |

(2) Accuracy (Acc)

Accuracy is the fraction of correctly classified data points in the total dataset. In this case, we use maximum matching to find out the correspondence between $P$ and $C$.

Given a vertex $v_i$, $l_{ci}$ is the class label of $v_i$ assigned by an algorithm and $l_{pi}$ is its true label. The Accuracy is defined as

$$\text{Accuracy} = \frac{1}{n}\sum_{i=1}^{n} \delta(l_{pi}, f_{map}(l_{ci})) \quad (5)$$

where $\delta(\cdot)$ is a Kronecker function which returns 1 if the two variables are equal and 0 otherwise. $f_{map}(\cdot)$ is a mapping function that maps the label $l_{ci}$ to its corresponding maximum matching label $l_{pi}$ in the ground truth, $n$ is the total number of data point in the dataset.

## IV. EXPERIMENTS AND EVALUATION

In this section, we conduct a series of experiments to evaluate the features and our community detection method based on E-N algorithm. In subsection A, we make a comparison of two feature sets in terms of accuracy and efficiency in community detection. We discuss their advantages and limitations in detail. In subsection B, we compare E-N graph with epsilon graph and k-NN graph in terms of community detection performance and time complexity. To demonstrate the superiority of our community-based classification method over traditional clustering method, we also take k-means into comparison. In subsection C, we give an intuitional results of the community detection performance and the underlying relations between malapp families.

We collect 3996 malapp samples, which are from 13 popular malapp families, as our experiment dataset. The description of families can be seen in Table III.

In the feature extraction phase, we decompile APK files and extract 29967 features, including 810 platform-defined features and 29157 app-specific features.

The experiments are constructed on a PC with two quad-core 3.4 GHz i7 processors and 16G memory. Our method is implemented in Python language.

TABLE III MALAPP DATASET

|   | Family | Size | Brief Introduction |
|---|---|---|---|
| 1 | Opfake | 1004 | Send premium SMS messages |
| 2 | Plankton | 868 | Communicate with remote servers, download and install other applications, send premium SMS messages |
| 3 | Youmi | 329 | An advertisement library that compromises personal information |
| 4 | Utchi | 305 | An advertisement library that compromises personal information |
| 5 | Basebridge | 303 | Forward confidential details (SMS, IMSI, IMEI) to a remote server |
| 6 | Kmin | 248 | Send Android device data to a remote server |
| 7 | Iconosys | 153 | Steal personal data |
| 8 | Mseg | 188 | Steal private data and secretly send SMS to premium-rated numbers |
| 9 | Fakedoc | 147 | Install additional applications |
| 10 | Appquanta | 140 | Adware |
| 11 | Geinimi | 108 | Open a back door and transmit information from the device (IMEI, IMSI, etc.) to a specific URL |
| 12 | Gingerbreak | 103 | A root tool |
| 13 | Smsspy | 100 | Banking Trojan targeting consumers in Spain |
| Total |  | 3996 |  |

*A. Feature Comparison*

The static features we extracted can be divided into platform-defined features and app-specific features. Platform-defined features are defined by Android system, e.g., the hardware and system APIs. App-specific features are generated by developers, e.g., strings such as IP or URL in the hard code. Since the app-specific feature set is much larger than platform-defined feature set, it is necessary to figure out whether the expense is worthwhile.

We detect communities based on all features and only platform-defined features respectively. The detection performance is shown in Fig. 3. The y-axis represents the measurements of rand statistic or accuracy. The x-axis represents values of $p$. $p$ is a set by intervals, meaning that we create an edge for $v_i$ and $v_j$ if $w_{ij}$ is among the top $p\%$ of weight set. E.g., when $p$ is 10, edges will be built for vertex pairs whose weights are among the top 10% of $W$.

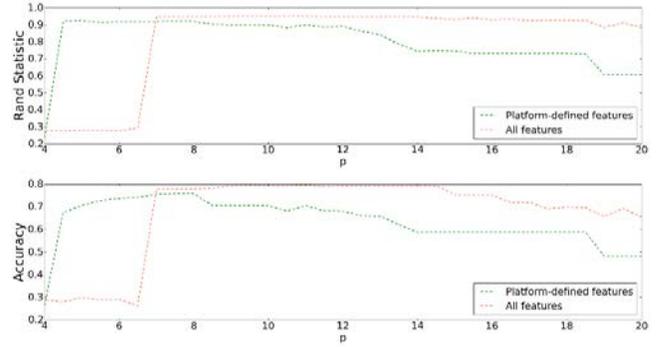

Fig. 3. Community detection performance based on two feature sets

TABLE IV COMMUNITY DETECTION PERFORMANCE BASED ON TWO FEATURE SETS

|  | Platform-defined Features | | All Features | |
|---|---|---|---|---|
| $p$ | RS (%) | Acc (%) | RS(%) | Acc(%) |
| 20 | 60.65 | 48.2 | 88.41 | 65.49 |
| 9.5 | 90.06 | 70.47 | 94.93 | 79.53 |
| 7.5 | 92.02 | 75.8 | 94.69 | 77.58 |
| 4 | 22.65 | 25.9 | 27.5 | 28.85 |

From Fig. 3 we have two observations.

(1) Both feature sets' performance shows a tendency of rise-steady-descend. If $p$ is set to a small value, which means that the corresponding $\varepsilon$ is large, only the most similar vertex pairs have edges. The sparse edge set is unable to contain enough information to form communities, which leads to low detection accuracy. On the contrary, when $p$ is large, two vertices will be linked together even if they are not so similar. The numerous edges lead to obscure community boundaries. For instance, when $p$ is 20, the average degree of a vertex is 400. When $p$ is 5, the degree drops to 101. When $p$ is set around 10, both feature sets achieve best performance.

(2) In general, the set of all features achieves better accuracy than platform-defined features. The set of all features contain both platform-defined features and app-specific features. It confirms the effectiveness of signatures on profiling malapps and distinguishing known families. As shown in Table IV, when $p$ is 9.5, the set of all features achieves best performance, with RS of 94.93% and accuracy of 79.53%. However, with only platform-defined features, we can achieve best accuracy of 75.8%, only 4% lower than using all features, but reducing the size of feature set from 29157 to 810. It reveals that our approach is able to be independent of signatures or any prior patterns and has better practicability, which is a requirement to detect novel malapps.

According to our previous work [5], though with app-specific features we can achieve better performance, platform-defined features are more persistent. In addition, in terms of quantity, the app-specific features grow with the increase of dataset, while the number of platform-defined features keeps stable instead. We rank the platform-defined features by their frequencies and obtain the top frequently used features. From Table V, we observe that 98.32% malapps ask for the Internet permission and 59.1% tend to send messages. 71.42% malapps invoke "getDeviceId" API to get the unique identifier of the devices.

TABLE V TOP 12 PLATFORM-DEFINED FEATURES IN ALL FAMILIES

|   | Feature Name | Percentage | Feature Set |
|---|---|---|---|
| 1 | Internet | 98.32% | Permission |
| 2 | Read_Phone_State | 97.52% | |
| 3 | Access_Network_State | 84.64% | |
| 4 | Write_External_Storage | 78.30% | |
| 5 | Receive_Boot_Completed | 71.45% | |
| 6 | Send_Sms | 59.10% | |
| 7 | Read_Sms | 46.32% | |
| 8 | Reflection Used | 82.76% | Code Related |
| 9 | Get System Service | 71.42% | Restricted API |
| 10 | Get Active Network Info | 65.62% | |
| 11 | Telephony | 64.59% | Hardware |
| 12 | Location | 42.34% | |

TABLE VI TOP 5 MOST COMMONLY INVOKED SENSITIVE APIS IN DIFFERENT FAMILIES

| Family | Sensitive API |
|---|---|
| Opfake | Landroid/telephony/TelephonyManager;->getDeviceId |
| | Landroid/telephony/TelephonyManager;->getLine1Number |
| | Landroid/telephony/SmsManager;->sendTextMessage |
| | Landroid/telephony/gsm/SmsManager;->sendTextMessage |
| | Landroid/webkit/WebView;->loadUrl |
| Fakedoc | Landroid/app/ActivityManager;->restartPackage |
| | Landroid/net/ConnectivityManager;->getActiveNetworkInfo |
| | Landroid/telephony/TelephonyManager;->getDeviceId |
| | Landroid/net/ConnectivityManager;->getNetworkInfo |
| | Landroid/net/wifi/WifiManager;->setWifiEnabled |
| Ginger break | Landroid/telephony/TelephonyManager;->getDeviceId |
| | Ljava/net/HttpURLConnection;->connect |
| | Landroid/net/ConnectivityManager;->getActiveNetworkInfo |
| | Landroid/app/ActivityManager;->getRunningTasks |
| | Landroid/location/LocationManager;->getLastKnownLocation |

Malapps implement malicious operations by invoking sensitive APIs. APIs related to device information, network and location are most frequently used in malapps. Different families tend to have different rankings in APIs. We take Opfake, Fakedoc and Gingerbreak as an example. Table VI shows the top 5 sensitive API sets of these three families. Opfake's malicious behavior is to send premium SMS messages. "SendTextMessage" is its 3rd commonly used sensitive API. Fakedoc aims to install additional applications. Thus establishing network connection and manipulating applications are its primary requirements. Gingerbreak is a tool to get the root permissions of the phones. It accesses devices' information and users' locations by "getDeviceId" and "getLastKnownLocation." It monitors other applications by invoking "getRunningTasks" to get the activity which is being displayed. The same feature has different ranks in different families can ensure tf-idf achieves maximum effectiveness.

### B. Evaluation of E-N Method

In this subsection, we employ the E-N method to construct the relation graph and test its performance on community detection. In order to facilitate comparisons with existing graph construction methods, we also conduct experiments on k-NN graph and epsilon graph. Moreover, to demonstrate the superiority of our community-based classification method over traditional clustering method, we took a comparison with the popularly used clustering algorithm, k-means [40].

As shown in Fig. 4, k-NN graph is sensitive to the parameter $k$. When $k < 175$, communities cannot be clearly found in the graph. When $k \in [180, 290]$, it achieves the best detection accuracy of 74.97%.

Epsilon graph and E-N graph show a tendency of rise-steady-descend. The reason has been discussed in subsection A. Picking the appropriate $p$ is much easier than $k$ since $k$ is in a wide range of $[1, n]$. Table VII and VIII show the detailed results. With platform-defined features, E-N graph achieves RS of 92.42% and accuracy of 77.73%. With all features, it has the best performance with RS of 94.93% and accuracy of 79.58%. In general, E-N graph achieves better performance than the other two graphs, which confirms the efficiency of the E-N method.

Work exists [41-43] on categorizing malapps that applied the k-means algorithm, however, the number of clusters need to be manually set. Given a set of malapps, it is difficult to tell the real number of their belonging families. In addition, k-means shows bad performance on data of which the hidden class is highly imbalanced. It is also very sensitive to outliers and noise. In our experiments, k-means fail to cluster the malapps. It only achieves the best accuracy of 41.99%.

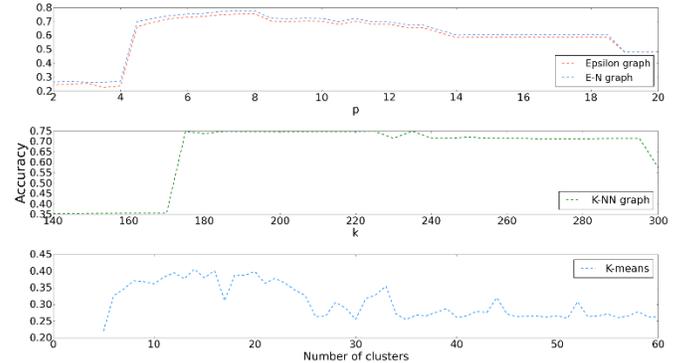

Fig. 4. Community detection performance based on k-NN graph, epsilon graph and E-N graph. The figure on the bottom shows the performance of k-means.

TABLE VII PERFORMANCE OF MALAPP CLASSIFICATION BASED ON DIFFERENT METHODS WITH PLATFORM-DEFINED FEATURES

| Method | Settings | RS(%) | Acc(%) |
|---|---|---|---|
| Epsilon Graph | $p$=8 | 92.12 | 75.61 |
| K-NN Graph | $k$=235 | 92.68 | 74.97 |
| E-N Graph | $p$=8, $k$=1 | 92.42 | 77.73 |
| K-means | $c$=14 | 84.20 | 40.57 |

TABLE VIII PERFORMANCE OF MALAPP CLASSIFICATION BASED ON DIFFERENT METHODS WITH ALL FEATURES

| Method | Settings | RS(%) | Acc(%) |
|---|---|---|---|
| Epsilon Graph | $p$=11 | 94.93 | 79.53 |
| K-NN Graph | $k$=235 | 92.68 | 74.97 |
| E-N Graph | $p$=11, $k$=1 | 94.93 | 79.58 |
| K-means | $c$=17 | 84.76 | 41.99 |

To evaluate the E-N method in terms of time consumption, we compare the execution time of the three graph construction methods with different size of dataset, which is shown in Fig. 5. K-NN suffers from high time complexity. With the number of vertices increasing, its execution time rises rapidly. The epsilon graph achieves the best performance in terms of time complexity. In E-N algorithm, most vertices only need the first step. The number of isolated vertices which are left to the k-NN phase is limited. Hence the execution time does not increase distinctly, compared to original epsilon method. When the number of vertices researches 1000, it only takes 35 seconds to build the relation graph with E-N method. For the whole 3996 features, it only needs 10 minutes.

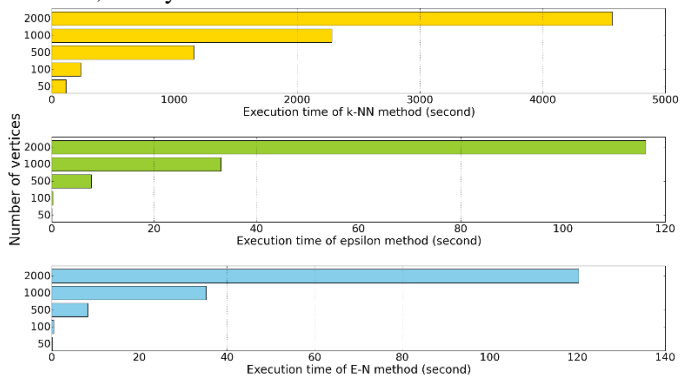

Fig. 5. The execution time of graph construction based on three methods

### C. Performance of Community Detection

In the graph construction step, we have built a relation graph based on our dateset. We can obscurely observe the relations among malapps in Fig. 6. Some malapps are densely connected, while some are dispersedly distributed. In order to extract the underlying relations in the graph, obtain distinct boundaries and detect communities, we apply Infomap. As shown in Fig. 7, the relation graph is divided into several dependent communities and the structures can be clearly observed.

Fig. 8 is a matrix describing the E-N graph based community detection results of different families. The y-axis represents 13 test malapp families and the x-axis represents 27 communities found in the relation graph. The color block represents the percent of elements in $x$ family assigned to $y$ community. The more malapps in the corresponding community, the brighter the color is. For Appquanta, Basebridge, Fakedoc, Geinimi, Iconosys, Mesg, Plankton and Utchi, the blocks' colors are bright yellow, which means that almost all the malapps in these eight families are clustered into corresponding eight communities distinctly. E.g., 100% samples in Appquanta have been assigned to community X5, 99. 65% samples in plankton are in community X1. However, a few mismatches exist. 25% samples in Smsspy and 21.3% samples in Youmi are both assigned to X12 community. We found that 42.4% features in Smsspy appear in Youmi. They are joint together since weights between these samples are above the threshold.

To get an intuitional observation of the potential relations between families, we calculate the weights for each family pair. As shown in Fig. 9, the circles represent the weights among families. If two families have similar behaviors as well as high weight, the circle will be big and bright. We find that the high score may appear between families that don't have similar descriptions. E.g., Smsspy is known as a banking Trojan family and Youmi is an adware family. From the descriptions, they seem to be unlike. However, as previously discussed, they have lots of common features and part of Smsspy and Youmi are divided into the same community. They share a weight of 3.8 while the weight between Smsspy and another family, Appquanta, is only 1.4.

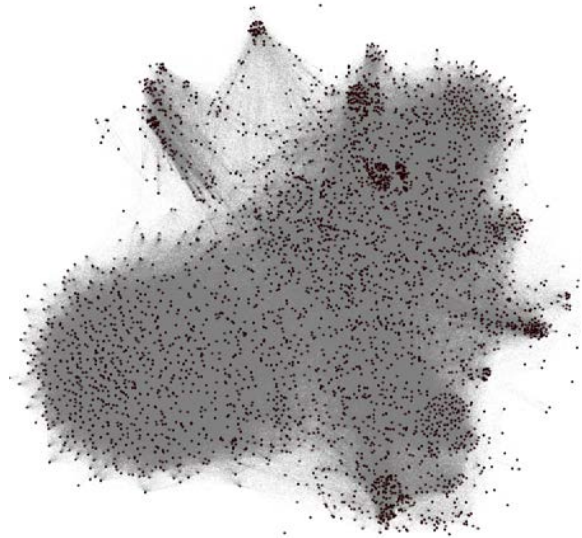

Fig. 6. The relation graph of malapps

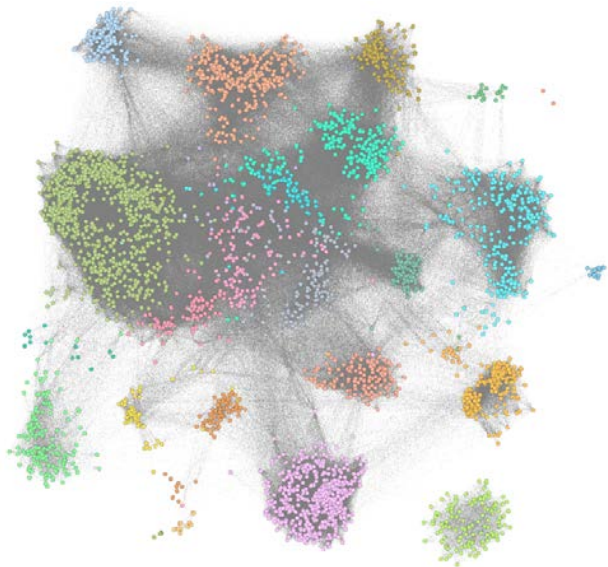

Fig. 7. Communities of malapps

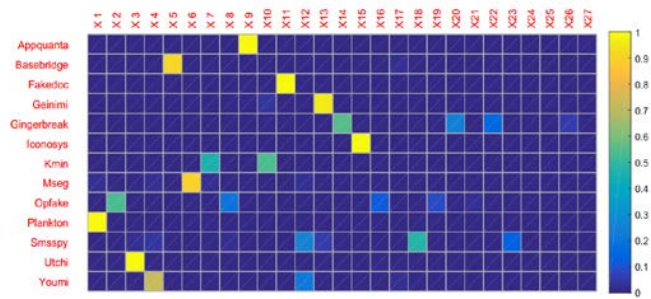

Fig. 8. Community detection results of different families

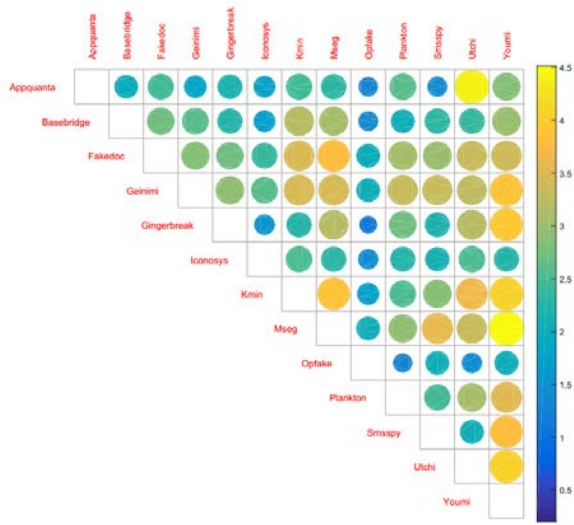

Fig. 9. Similarities between malapp families

*D. Discussion*

We build a relation graph and utilize a community detection method to extract relations among malapps, which achieves better results than traditional clustering methods and reduces the time consumption in the same time. Community detection methods have been widely used in social network. We notice that the relations between people and communities are much alike the relations between malapps and families. Therefore, we test the community detection method on malapp samples and the results prove it. In comparison, traditional clustering methods like k-means fail to cluster families of the multiple samples.

The advantage of graph construction and community detection methods over traditional clustering methods, e.g., k-means, mainly lies in the ability to explore the implicit similarities of interrelated data. Clustering methods regard vertices as independent. But from the perspective of graph, the vertices are interrelated. By building a relation graph, we can describe the similarities between vertices by common feature sets directly and also by common nearest neighbors implicitly. When two malapps have no common features, traditional clustering methods would take them as different. But community detection methods can still capture the similarities between them through the edges of their neighbors. On the other hand, the features we observe may not be complete and precise. Community detection methods have better adaptability to noise and outliers because of the additional similarity information provided by the graph edges.

Though we utilize community detection to find malapp families, there exists some differences. Both communities and families are clusters of malapps that have similar behaviors. Specifically, malapp families are classified with manual analyses and interventions. However, some similarities between malapps are not directly provided, but hidden in the edges in the relation graph. The community detection is an unsupervised as well as autonomous process. It can automatically discover the relations without manual work. Given malapp samples, communities can be obtained initiatively without any prior knowledge.

*E. Limitation*

The limitation is that our community detection based method for malapp family classification is not as precise as supervised learning approaches. However, it doesn't rely on signatures or any prior learned patterns, and is able to detect novel malapps. It gives explainable results of the distribution of communities. Moreover, it shows better ability to cope with the situations of sample shortage and abundant unlabeled malapp samples. Faced with unknown malapps, it can automatically identify the quantity of families which is hard to determine for many other methods.

V. CONCLUSION

In response to the current trend of rapidly increasing number of malapp variants on Android-based MCPS, we propose a malapp community discovery framework to uncover the underlying relations among malapps. 11 categories of features are extracted to profile the behaviors of malapps in different aspects. Based on these behaviors, a relation graph is built to reveal the similarities among malapps. For the purpose of overcoming the disadvantages of epsilon graph and k-NN graph, we propose the E-N method in the relation graph construction step. The experimental results demonstrate its effectiveness and efficiency. We employ community detection technique to discover underlying relations of malapps in the relation graph. It has better ability to explore implicit similarities of interrelated data, compared with traditional clustering method. The experimental results verify the superiority of our method. It achieves the best performance with rand static of 94.93% and accuracy of 79.53%.

In the future, we plan to explore more effective graph construction algorithms to better profile relations among malapps on Android-based MCPS. More features to characterize malapps' behaviors are also being investigated.